\documentclass[journal]{IEEEtran}
\usepackage[T1]{fontenc}
\usepackage{graphicx,colordvi,psfrag}
\usepackage{amsmath,amssymb,algpseudocode}
\usepackage{algorithm}
\usepackage{algorithmicx}
\usepackage{calc,pstricks, pgf, xcolor}
\usepackage{bbding}
\usepackage{bbm}
\usepackage{dsfont}
\usepackage{stfloats}
\usepackage{hyperref}
\usepackage{xparse}
\usepackage{enumerate}
\usepackage{tikz}
\usetikzlibrary{shapes,arrows,positioning,calc}

\usepackage{subcaption}

\newcommand{\Ind}{\mathds{1}}

\newcommand{\ZZ}{\mathbb{Z}}
\newcommand{\RR}{\mathbb{R}}

\newcommand{\Var}{\mathrm{Var}}

\newcommand{\eps}{\varepsilon}

\newcommand{\Unif}{\mathrm{Uniform}}
\newcommand\indep{\protect\mathpalette{\protect\independenT}{\perp}}
\def\independenT#1#2{\mathrel{\rlap{$#1#2$}\mkern2mu{#1#2}}}

\def\eqdef{\triangleq}

\def\EE{\mathbb{E}}
\newcommand{\m}{\mathcal}

\newcommand{\round}{\mathrm{round}}
\newcommand{\Reff}{R_\mathrm{eff}}

\newcommand{\covol}{\mathrm{covol}}
\newcommand{\ERho}{C_{\mathrm{FP}}}

\NewDocumentCommand{\DIP}{e{^_}}{D^{\mathrm{IP}\IfValueT{#1}{,#1}}_{\IfValueT{#2}{#2}}}

\NewDocumentCommand{\DMM}{e{^_}}{D^{\mathrm{MM}\IfValueT{#1}{,#1}}_{\IfValueT{#2}{#2}}}

\newcommand{\simiid}{\stackrel{\mathrm{iid}}{\sim}} 

\def\dperp{\perp\!\!\!\perp}
\def\calS{\mathcal{S}}

\DeclareMathOperator*{\argmin}{\arg\!\min}

\def\eqdef{\triangleq}

\begin{document}
\title{High-Rate Quantized Matrix Multiplication I}

\author{Or Ordentlich 
and	Yury Polyanskiy 
\thanks{O. Ordentlich is with the 
Hebrew University of Jerusalem, Israel (\texttt{or.ordentlich@mail.huji.ac.il}).  Y. Polyanskiy is with the MIT,
USA (\texttt{yp@mit.edu}). The work of OO was supported by the Israel Science 
Foundation (ISF),
grant No. 2878/25. The work of YP was supported in part by the MIT-IBM Watson AI Lab and by the National Science
    Foundation under Grant No CCF-2131115. }
}

\date{}

\maketitle
\begin{abstract}

This paper investigates the problem of quantized matrix multiplication (MatMul), which has become
crucial for the efficient deployment of large language models (LLMs). We consider a
Generic MatMul setting, where both matrices must be quantized (weight+activation quantization)
without specific apriori (calibration) statistical information about the factors. We review the fundamental information-theoretic tradeoff
between quantization rate and distortion (high-rate theory), and contrast those with the performance of
popular quantization schemes (absmax INT and floating-point (FP)), 
for which we also derive accurate heuristic approximations. Part II of this paper studies the weight-only quantization setup where second-order statistics of the activation matrices are available at the encoder.
\end{abstract}


\section{Introduction}

Matrix multiplication (MatMul) is the work-horse of AI. Consequently, great amounts of research efforts are invested in exploring ways in which MatMul can be executed with as few resources as possible, while maintaining satisfactory accuracy for the underlying application.

One of the most prominent techniques for achieving this goal is quantization, where full-precision matrices are first compressed to $R$ bits per entry, and MatMul is computed based on the compressed descriptions of the matrices. In LLMs, quantization serves two main purposes: 1)Reducing the IO burden, which is often the bottleneck for MatMul. This is achieved since the number of bytes that needs to be moved around the hardware scales linearly with $R$; 2)Representing the matrices with small data-types, e.g., INT8, FP8, INT4, FP4, such that on top of IO savings one can also accelerate compute by using faster multipliers. 

In light of the above, the last few years have seen an explosion in the number of papers published on quantized matrix multiplication for LLMs, mostly in the AI and machine learning literature. However, the fundamental information theoretic limits on the tradeoff between quantization rate and distortion in quantized MatMul were only recently established~\cite{ordentlich2024optimal}, and only for the case of generic MatMul, where we have no prior knowledge on the characteristics of the matrices to be multiplied when designing the quantizers.

The purpose of this paper, which is the first-part of a two-part paper, is to put forth information theoretic benchmarks for
generic quantized MatMul, and use them for evaluating the quality of several popular quantization
schemes, and their gap to the fundamental limit. Part II of this paper~\cite{OP26part2} studies weight-only quantization, where the activations are given in full-resolution and their second-order statistics can be used when quantizing the weights matrices.


We stress from the outset a critical distinction in the meaning of the word ``rate'' between
information theorists and practicioners. The former, focusing on fundamental limits, assume that
quantization algorithm inputs a vector/matrix of $N$ entries and outputs $NR$ bits, where each bit
is dependent on all of the input entries (e.g., as in vector quantization). For the latter,
quantization algorithm takes each entry, applies a simple rule  to each entry and produces one of
$2^R$ possible discrete outputs (e.g., as in standard INT$M$ or FP$M$ formats). Recently, so called
\textit{microscaling} formats started offering a middle ground between the two: a simple scaling
operation is applied to a small group of 16 or 32 entries, after which each entry is still
processed individually (e.g., as in NVFP4 and MXFP4). The purpose of this survey is to quantify
the gap between fundamental limits and the restrictive practical approaches, specifically in the
context of LLM quantization.

In order to enable clean and simple analysis, in this paper we restrict attention to high-resolution quantization. This essentially boils down to assuming that the strength of the quantization noise is significantly weaker than that of the signal, which in turn, enables to neglect several terms in the distortion analysis. While the high-resolution assumption may lead to erroneous conclusions for low rate, say $R<2$ bits per entry, for relatively high-quantization rate it usually provides sufficiently accurate expressions for the purpose of this survey paper, with the advantage of simplified analysis.

\subsection{Organization and summary of results}

Below we give a short overview of the content of this part of the survey. In
Section~\ref{sec:matmul_theory} we consider the problem of rate-$R$ quantized ``generic'' MatMul,
where no a priori knowledge of the matrices is available when designing the quantizers. We give an
overview of known theoretic upper and lower bounds and conclude that in this setup the smallest
expected distortion for the $ij$th entry of the matrix product is $K(i,j)\cdot 2\cdot 2^{-2R}$,
where $K(i,j)$ is the product of squared $\ell_2$ norms of the vectors participating in the
corresponding inner product, normalized by their dimension. 

In Section~\ref{sec:matmul_practice} we analyze the performance of several popular MatMul
quantization schemes including INT multipliers with absmax scaling, floating-point (FP)
multipliers with absmax scaling and NVFP4, as well as the recently proposed
NestQuant~\cite{NestQuant} scheme. For INT and FP multipliers we develop approximations on the
attained distortion of the form $K(i,j)\cdot2\cdot 2^{-2\Reff}$, from which their gap to
optimality in bits is immediately seen to be $R-\Reff$, where $R$ is the actual number of bits per
entry these schemes use. The FP analysis appears to be new and is empirically shown to be quite
accurate. As an example, we show that for FP we get $\Reff \approx \mathcal{M} + 2.23$ bit, where
$\mathcal{M}$ is the number of mantissa bits.

c

\subsection{High-rate assumption and uniform errors.}\label{sec:high-rate}

Some of the results in this survey (including
INT and FP analysis) work under the heuristic assumption that quantization can be modeled as
additive zero-mean, uniformly distributed error. Here we describe why this is accurate in the high-rate regime. 

\begin{figure*}[t] 
    \centering
    
    \begin{subfigure}[b]{0.48\textwidth}
        \centering
        \includegraphics[width=\linewidth]{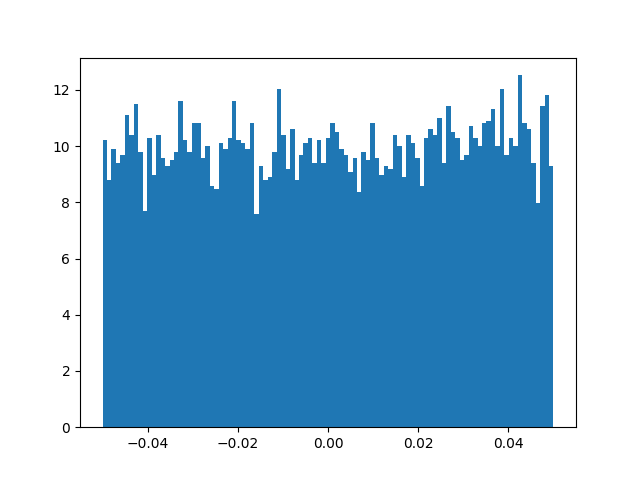}
        \caption{Histogram of rounding errors}
        \label{fig:rnd1}
    \end{subfigure}
    \hfill 
    \begin{subfigure}[b]{0.48\textwidth}
        \centering
        \includegraphics[width=\linewidth]{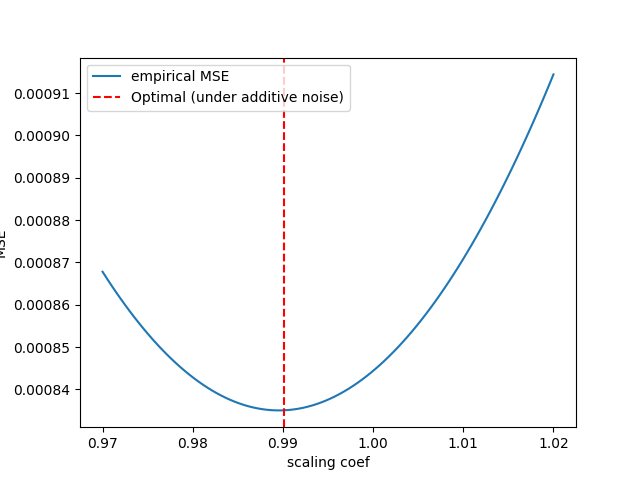}
        \caption{MSE as a function of scaling coefficient $\gamma$}
        \label{fig:rnd2}
    \end{subfigure}
    
    \caption{Distribution of $\epsilon$-quantization errors of a vector $\mathcal{N}(0,\sigma^2
    I_{10000})$ with $\sigma^2 = 1/12$ and $\epsilon=0.1$ and effects of shrinkage on MSE.}
    \label{fig:rnd}
\end{figure*}

As an example, consider a high-dimensional vector $x=(x_1,\ldots,x_n)$. 
Its approximation over an $\epsilon$-grid (quantization) is defined as $\tilde x_i = \epsilon\cdot
\mathrm{round}(x_i/\epsilon)$. The errors $e_i = \tilde x_i - x_i$ clearly are deterministic functions of $x_i$ and
should not be modeled as random. However, when $\epsilon$ is small and entries of $x$ are not ``in
any special position'' with respect to $\epsilon$-grid, we expect the residuals to land fairly
uniformly over the interval $[-\epsilon/2, \epsilon/2)$. (This can be easily checked empirically,
by sampling $x_i \simiid P$ from any unit-variance distribution $P$ with smooth
density and taking $\epsilon \ll 1$; see Fig.~\ref{fig:rnd1}.) Furthermore, generally the empirical average ${1\over n}
\sum_i x_i e_i \approx {1\over n} \sum_i \tilde x_i e_i \approx 0$,
cf.~\cite{bennett1948spectra,gray2002quantization}.\footnote{Recall also that our interest is in
the quality of approximating the inner product of two vectors $x,y\in\RR^n$ from the inner product
of their quantized versions $\tilde{x},\tilde{y}$. The error in approximating this inner product
is a scalar that depends on the $2n$ per-coordinate quantization errors, and therefore it is
indeed the ``average'' behavior of the quantization errors along the $n$ coordinates that will
dictate the performance. For the similar reason, inner-product quantization error, being a
quadratic function of $2n$ random variables (with linear part dominant in the high-rate regime),
will be normally distributed regardless of the precise statistics of each entry's quantization
noise.}

Consequently, under the $\epsilon \ll 1$ assumption we
can model the effect of quantization as passing data through an additive uniform noise channel:
\begin{equation}\label{eq:quant_channel}
	\tilde x_i = x_i + e_i,~~~ e_i \simiid \mathrm{Unif}[-\epsilon/2,\epsilon/2)\,.
\end{equation}
(This can be made rigorous by replacing uniform quantization with dithered uniform
quantization~\cite{ramiBook}. Alternatively, in some of the schemes described below, a random rotation is applied prior to quantization in order to ``Gaussianize'' the input to the quantizer. It is shown in~\cite{hadad2016dithered} that a by-product of such rotation is that the quantization errors become uncorrelated
as the dimension increases.)

Note that generally, our entries will be assumed to be of zero mean
and $O(1)$ (empirical) variance and hence $\tilde x_i$'s range over $O(1/\epsilon)$ possible
values. Thus, we can think of relationship between $\epsilon$ and rate as $\epsilon \asymp
2^{-R}$. Overall, the high-rate assumption allows us to model quantization error as stochastic, independent
uniform and additive. 

Another way in which high-rate (low $\epsilon$) assumption is helping us is in ignoring so-called
``linear shrinkage'' factors. Specifically, it turns out that despite $\tilde x$ being composed of
nearest elements of the $\epsilon$-grid to $x$, the best estimate of the vector $x$ given $\tilde
x$ is NOT $\tilde x$ itself. I.e. one could benefit from setting
$$ \hat x = \gamma \tilde x\,,$$
for some $\gamma \le 1$. To justify this, assume that~\eqref{eq:quant_channel} is the correct
model. Then 
\begin{align*} {1\over n}\|x - \hat x\|_2^2 &= {1\over n} \sum_i ((1-\gamma) x_i + \gamma e_i)^2
\\
	&\approx (1-\gamma)^2
\hat {\mathbb{E}} [X^2] + \gamma^2 {\epsilon^2\over 12}\,.
\end{align*}
Optimizing this over $\gamma$ gives 
$$ \gamma^* \approx {\hat {\mathbb{E}} [X^2] \over \hat {\mathbb{E}} [X^2] + {\epsilon^2\over 12}} = 1 -
O(\epsilon^2)\,,$$
where $\hat {\mathbb{E}}[X^2] = {1\over n} \sum_i x_i^2$ is the empirical second moment.
This somewhat surprising \textit{shrinkage effect} is empirically demonstrated on
Fig.~\ref{fig:rnd2}. We do want to emphasize that the above estimate of $\gamma^*$ is only
an approximation of the optimal shrinkage factor. For example, if $x_i \simiid
\mathrm{Unif}[-{1+\epsilon\over 2},{1+\epsilon\over 2})$ then the optimal shrinkage factor is
exactly $\gamma^*=1$ despite the additive-noise estimate giving a value of around $1-\epsilon^2$.

Overall, the MSE improvements from applying shrinkage are on the order of $O(\epsilon^4) =
O(2^{-4R})$. Thus, the second impact of the high-rate assumption is that we will be ignoring
effects of shrinkage, and also dropping all terms of order below $2^{-2R}$ from discussion of
fundamental limits.

\section{Quantized matrix multiplication: Theory}\label{sec:matmul_theory}

Let us introduce the problem formally. Consider two matrices $A\in\RR^{n \times a}$ and
$B\in\RR^{n\times b}$. Our goal is to compute their product $A^\top B$, which is simply a matrix
of $ab$ inner products of columns of $A$ and $B$. The problem is that the computation device is
only able to receive information about matrices $A$ and $B$ with a constrained rate of $R$ bits
per entry. Because of this limitation, computing exact product $A^\top B$ is not possible and hence we need to
accept a certain error. The goal of quantized matrix multiplication is to reduce this error. 

More formally, a rate $R$ quantization scheme $\calS$ for matrix multiplication $A^\top B$ where
$A\in\RR^{n \times a}$ and $B\in\RR^{n\times b}$ consists of:
\begin{enumerate}
\item A random variable $\omega \in \Omega$ (shared common randomness, or ``public coin'')
\item A pair of encoders $f_1:\RR^{n \times a}\times \Omega \to[2^{naR}]$, $f_2:\RR^{n \times
b}\times \Omega \to[2^{nbR}]$. These encoders produce rate $R$ descriptions of two matrices.
\item A decoder $g:[2^{naR}]\times [2^{nbR}]\times \Omega\to\RR^{a\times b}$, which produces an
estimate of the matrix product. We will simply write $\widehat{A^\top B}=g(f_1(A, \omega),f_2(B,
\omega), \omega)$ to denote this approximation.
\end{enumerate}
The distortion of the scheme on the instance $A,B$ is defined as 
$$
D(A,B; \calS)=\frac{1}{nab}\EE_{\omega \sim P_\omega}[\|A^\top B-\widehat{A^\top B}\|_F^2]\,.
$$
Note that we normalize the squared Frobenius norm of the error matrix by $nab$. The $ab$ term accounts for the number of entries in the error matrix, while $n$ accounts for the typical squared magnitude of each entry in $A^\top B$. Indeed, each such entry is the inner product between two vectors in $\RR^n$. If those entries are drawn from a fixed iid distribution, the variance of the inner product will be $O(n)$, and for fixed quantization rate $R$ the squared quantization error will also be $O(n)$.

To complete the problem setting we need to agree on what is known a priori about matrices
$A$ and $B$. We will consider two cases: worst-case (where nothing, except norm constraints, are
assumed about $A$ and $B$) and average-Gaussian.

\subsection{Quantization: worst-case setting}

Let us consider the worst-case performance of a scheme, defined as 
$$ D(\calS) = \sup_{A,B} D(A,B;\calS)\,,$$
with supremum over matrices $A$, $B$ with all columns of $\ell_2$ norm $\sqrt{n}$. Throughout this
subsection we will assume this restriction on $A,B$ unless noted otherwise. 

The first and most natural scheme is to quantize each column of $A$ and $B$ using a rate-$R$
vector quantizer over the sphere $\sqrt{n}\mathbb{S}^{n-1}$ and then set $\widehat{A^\top B} =
\hat A^\top \hat B$. It is well known that the best spherical
quantizer is able to achieve approximation guarantee (uniformly over all $A$) s.t.
	$$ \|A-\hat A\|_F^2 \le \|A\|_F^2 2^{-2R+o(1)} = na 2^{-2R+o(1)}\,.$$
How do we bound worst-case distortion? Suppose that we were lucky and $\hat B = B$, then the best
bound we can have is from Cauchy-Schwarz
$$ \|\hat A^\top \hat B - A^\top B\|_F^2 = \|B^\top (\hat A - A)\|_F^2 \le \|B\|_F^2 \|\hat A -
A\|_F^2\,,$$
which is tight for the case when $B$ and the quantization error $\hat A - A$ are co-aligned.
Dropping the assumption of $\hat B=B$ and applying triangle inequality one can show using the same
method the bound 
$$ D(\calS) \le n 2 \cdot 2^{-2R + o(1)}\,.$$
This is not good since we were expecting to get bounded distortion (independent of $n$).

Can this
adversarial alignment between $B$ and quantization error of $A$ be prevented by a better scheme?
The answer is negative, at least for \textit{any scheme that does not use common
randomness $\omega$}. Specifically, ~\cite[Proposition 1]{ordentlich2024optimal} shows that any
such non-randomized quantizer scheme $\calS$ for any finite rate $R>0$ must satisfy 
$$ D(\calS) \ge \delta(R) n\,,$$
where $\delta(R)>0$ and is independent of $\calS$.

The next idea is that of sketching~\cite{alon1996space}, which is often used in the literature on
approximate matrix multiplication. For simplicity, let us further assume $a=b=1$. Then define
$U\sim \mathcal{N}(0,I_n)$ and let it be shared between the encoders $f_1$ and $f_2$. 
Let us then compute two scalar quantities $U_A\eqdef U^\top A$ and
$U_B \eqdef U^\top B$. Notice that 
$$ \EE[U_A U_B] = A^\top \EE[ U U^\top] B = A^\top B\,,$$
and hence the product $U_A U_B$ yields an unbiased estimate of the inner product. Now, since these
two scalar quantities are with high probability bounded by $O(\sqrt{n})$ they can be easily
quantized with exponentially small mean squared error given the budget of $nR$ bits. Unfortunately, this idea
also does not work, as a simple computation shows
\begin{align*} \Var[U_A U_B] &= \EE[U_A^2 U_B^2] - (A^\top B)^2 \\&= n^2 \EE[U_1^2 (\rho U_1 + \sqrt{1-\rho^2}U_2)^2]
- n^2 \rho^2 \\&= n^2(1+\rho^2)\,,
\end{align*}
where $\rho = {1\over n} A^\top B$ is the cosine of the angle between $A$ and $B$. We see that
even if $U_A$ and $U_B$ were provided to the decoder exactly, the error would still be
$\Omega(n^2)$ due to variance of the estimate $U_A U_B$. Thus this kind of scheme again only
attains guarantee
$$ D(\calS) \gtrsim n\,. $$

Of course, one could reduce variance by
working with multiple sketches, however, to drop it all the way to $O(n)$ one would need order $n$
sketches and then $nR$ bits would need to be spent on quantizing an $n$ dimensional vector,
bringing us back to the original problem.

So far, we have seen that all non-randomized constructions as well as those based on sketching are
only yielding $\Omega(n)$ distortion. Nevertheless, by combining randomization (in the form of
random rotation and dither) with good lattice quantizers~\cite[Theorem 3]{ordentlich2024optimal}
constructs a scheme attaining guarantee
\begin{equation}\label{eq:det_lbnd}
	D(\calS) \le \frac{2\cdot 2^{2R}-1}{(2^{2R}-1)^2} + o(1)\,.
\end{equation}

In fact, the scheme achieves stronger guarantee even after dropping the assumption of normalized
columns. Specifically, for all matrices $A$ and $B$ with non-zero
columns of norms upper- and lower-bounded by polynomials in $n$, we have 
\begin{align}
\lefteqn{\forall i\in[a],j\in[b]~:~}\nonumber\\
\lefteqn{\EE_\omega \left((A^\top B)_{ij}-(\widehat{A^\top B})_{ij} \right)^2}& \nonumber\\
&\leq\frac{\|a_i\|^2\cdot \|b_j\|^2}{n}\left( \frac{2\cdot 2^{2R}-1}{(2^{2R}-1)^2}+o(1)\right)\nonumber\\
&\approx \frac{\|a_i\|^2\cdot \|b_j\|^2}{n} 2\cdot (2^{-2R}+o(1))\,.
\label{eq:Dachievable}
\end{align}
In the last step we applied our high-rate assumption and thus dropping terms smaller than
$2^{-2R}$ from consideration.

As will be seen from the next section, for Gaussian matrices $A$ and $B$ this upper bound is
essentially tight. Thus, this performance guarantee cannot be significantly improved (except
possibly by terms which are $o(2^{-2R})$ as $R\to \infty$).

Consequently, when designing rate $R$ quantization schemes for generic matrices, with $R\gg 1$, the smallest distortion we can attain simultaneously for all $A\in\RR^{n\times a}$, $B\in\RR^{n\times b}$ is characterized by~\eqref{eq:Dachievable}, and we therefore refer to
\begin{align}
D^*_{ij}=\frac{\|a_i\|^2\cdot \|b_j\|^2}{n} 2\cdot 2^{-2R}, ~~i\in[a]
,j\in[b]
\label{eq:fundlimit}
\end{align}
as the fundamental limit for high-resolution quantized generic matrix multiplication.

\subsection{Quantization: the iid Gaussian case}

So far we have only discussed the upper bounds. In order to prove lower bounds, we switch to
studying the random-input case. Specifically, we define for any quantization scheme $\calS$
\begin{align*} D_{\text{Gaussian}}(\calS) &= \EE_{A,B} D(A,B;\calS) \\&=  
\frac{1}{nab}\EE_{A,B,\omega}[\|A^\top B-\widehat{A^\top B}\|_F^2]\,,
\end{align*}
where $A_{i,j}\dperp B_{i,k} \simiid \mathcal{N}(0,\sigma^2)$.

\cite[Theorem 2]{ordentlich2024optimal} shows that for any scheme $\calS$ of rate $R$ we have
(non-asymptotically, for all $a,b,n$)
$$ D_{\text{Gaussian}}(\calS) \ge \sigma^4 \Gamma(R)\,,$$
where function $\Gamma(R)$ equals
\begin{align}
\begin{cases}
1-\left(1-\left( 2\cdot 2^{-2R^*}-2^{-4R^*}\right) \right) {R\over R^*}    & R\leq R^*\\
2\cdot 2^{-2R^*}-2^{-4R^*} & R>R^*
\end{cases},    
\label{eq:MatMulGaussRD}
\end{align}
where the critical rate is $R^*\approx0.906$.

Furthermore, it was shown in~\cite[Theorem 1]{ordentlich2024optimal} that there exists a
quantization scheme
$\calS$ (which is a variation of the lattice-based construction behind~\eqref{eq:Dachievable})
attaining
$$ D_{\text{Gaussian}}(\calS) \le \sigma^4 \Gamma(R) + o(1) $$
 for all $a,b$ and sufficiently large $n$. Therefore, the true fundamental limit was determined
 $$ \inf_{\calS} D_{\text{Gaussian}}(\calS) = \sigma^4 \Gamma(R) + o(1)\,.$$
 One interesting consequence of this result is demonstrating that (at least for Gaussian data and
 $R<R^*$) optimal strategy \emph{must} combine sketching, also known as Johnson-Lindenstrauss
 dimensionality reduction, with traditional vector quantization. Empirically, this
 combination algorithm was investigated in~\cite{zandieh2025qjl,zandieh2025turboquant}.

In this paper, as we agreed, we focus on the high-rate regime. In this regime, $R>R^*$ and it is
further assumed large enough such that $2^{-4R}\ll 2\cdot 2^{-2R}$ so that we can approximate
$\Gamma(R)\approx 2\cdot 2^{-2R}$. Overall, we see that on one hand, there exists a
scheme~\eqref{eq:Dachievable} that simultaneously attains
$$ \EE[\|\widehat{A^\top B} - A^\top B\|_F^2] \lesssim {\|A\|_F^2 \|B\|_F^2\over n} 2 \cdot
2^{-2R} $$
and that this is bound is not improvable generally, since for iid Gaussian $A,B$ we have
left-hand side lower bounded by the right-hand side (upto $o(2^{-2R})$ terms in the high-rate
regime). 

Thus, overall, we have theoretically backed reasons to think of $2 \cdot 2^{-2R}$ as fundamental
limit of quantized matrix product.

\section{Quantized generic matrix multiplication: Practice}\label{sec:matmul_practice}

In this section we restrict attention to inner product $x^\top y$ for $x,y\in\RR^n$. The matrix
product $A^\top B$ of $A=[a_1|\cdots|a_a]\in \RR^{a\times n}$ and $B=[b_1|\cdots|b_b]\in\RR^{b
\times n}$ consists of $a\times b$ inner products. In all schemes we discuss below, the
approximate MatMul is computed by separately quantizing each column of each matrix, and then
computing an approximation for each entry $(A^\top B)_{ij} \approx \hat a_i^\top \hat b_j$ by computing the inner product of 
quantized representations of $a_i\in\RR^n$
and $b_j\in\RR^n$. Thus, analyzing the distortion of these schemes for quantized inner product
immediately lends itself to analysis on the distortion of each entry in the matrix product.

We will assume throughout this section that per column scaling at perfect resolution is possible.
Mathematically, this means that $\hat x$ and $\hat y$ are represented in the form of $\hat x =
\gamma_x \tilde x$, $\hat y = \gamma_y \tilde y$, where $\gamma_x$ and $\gamma_y$ are
infinite-resolution scalars, and $\tilde x$, $\tilde y$ are high dimensional low precision
vectors.  In the AI literature this idea is sometimes called \emph{per-channel} and
\emph{per-token} scaling~\cite{xiao2023smoothquant}. In practice, these scalars are usually given in
FP16/FP32 resolution, which contributes a miniscule rate overhead due to high dimensionality of
the main vectors $\tilde x$, $\tilde y$. 


\subsection{INT$M$ Multipliers}

The constellation represented by an INT$M$ data-type is usually defined as
$\m{F}_{\mathrm{INT}M}=\ZZ\cap[-2^{{M-1}},2^{M-1})$. Here, to simplify expressions, we will extend
the constellation by one point and assume it consists of $\ZZ\cap[-2^{{M-1}},2^{M-1}]$. The most
straightforward way to approximate $x^\top y$ using INT$M$ multipliers is \emph{absmax
quantization} defined as follows:\footnote{We note that many openly available implementations of
absmax, e.g. derivatives of GPTQ~\cite{gptq-repo}, implement absmax suboptimally, so that the smallest integer value
of INT$M$ is either unused (50\% of time), or only used for one entry of the block (the other
50\%).}
\begin{itemize}
    \item Set $\gamma_x=2^{-(M-1)}\|x\|_{\infty}$, $\gamma_y=2^{-(M-1)}\|y\|_{\infty}$.
    \item Set $\hat{x}=\round(x/\gamma_x)$, $\hat{y}=\round(y/\gamma_y)$
    \item Set $\widehat{x^\top y}=\gamma_x\cdot\gamma_y\cdot \hat{x}^\top \hat{y}$.
\end{itemize}
Note that the definition of $\gamma_x,\gamma_y$ ensures that all entries of the integer vectors $\hat{x},\hat{y}$ are in $[-2^{M-1},2^{M-1}]$ and can therefore indeed be represented by the INT$M$ format. Let
\begin{align}
e_x=\hat{x}-x/\gamma_x,~~~ e_y=\hat{y}-y/\gamma_y.    
\end{align}
Following the discussion in Section~\ref{sec:high-rate}, for the analysis, we will make the simplifying assumption that
$e_x,e_y\stackrel{iid}{\sim}[-1/2,1/2)^n$ and are statistically independent of $(x,y)$. This
assumption is certainly incorrect, but nevertheless, it results in a simple and quite accurate
analysis for most pairs $x,y$ encountered in real-worlds applications. Furthermore, the assumption
can be made exact if one uses subtractive dithering~\cite{ramiBook}. 
We have
\begin{align}
D&_{\mathrm{INT}M}=\EE(\widehat{x^\top y}-x^\top y)^2\nonumber\\
&=\EE(\gamma_x e_x^\top y+ \gamma_y e_y^\top x+\gamma_x\gamma_y e_x^\top e_y)^2\nonumber\\
&=\frac{\gamma_x^2}{12}\|y\|^2 +   \frac{\gamma_y^2}{12}\|x\|^2+\frac{n\gamma_x^2\gamma_y^2}{12^2}\nonumber\\
&=\frac{\|x\|^2\cdot\|y\|^2}{n}\cdot\frac{2^{-2M}}{3}\nonumber\\
&\left[\frac{\|x\|^2_\infty}{\|x\|^2/n}+\frac{\|y\|^2_\infty}{\|y\|^2/n}+\frac{2^{-2M}}{3}\cdot\frac{\|x\|^2_\infty}{\|x\|^2/n}\cdot\frac{\|y\|^2_\infty}{\|y\|^2/n} \right].\nonumber
\end{align}
In what follows we will assume $M$ is large enough such that the last term above can be neglected
(this is manifestation of our focus on a high-resolution analysis). With this approximation, we
obtain 
\begin{align}
 D_{\mathrm{INT}M}\lesssim \frac{\|x\|^2\cdot\|y\|^2}{n} 2\cdot 2^{-2M}\cdot\frac{\Delta_{\mathrm{INT}}(x,y)}{3},
 \label{eq:DINTM}
\end{align}
where
\begin{align}
\Delta_{\mathrm{INT}}(x,y)=\frac{1}{2}\left( \frac{\|x\|^2_\infty}{\|x\|^2/n}+\frac{\|y\|^2_\infty}{\|y\|^2/n}\right).
\label{eq:DeltaINTdef}
\end{align}
We have that $\Delta_{\mathrm{INT}}(x,y)\leq n$ Since $\frac{\|x\|^2_\infty}{\|x\|^2/n}\leq n$,
and this is tight for the the natural basis vectors of $\RR^n$. Thus, for some matrices,
$D_{\mathrm{INT}M}$ exceeds the fundamental limit by a multiplicative factor of $\Omega(n)$. As we
have seen before, see~\eqref{eq:det_lbnd}, determinstic quantizers generally must suffer from
this kind of degradation.\footnote{Though, we note, here the situation is more subtle: the
quantization noise is assumed additive and independent of $x$, thus making quantizer
non-deterministic. But the noise is nevertheless correlated with $x$ through scaling factor.} 

A popular way of
preventing the situation of large $\Delta_{\mathrm{INT}}(x,y)$ is applying a random orthogonal
matrix $S\in \RR^{n\times n}$, drawn from the Haar distribution on the orthogonal group $\mathcal{O}(n)$, to both vectors prior to quantization, such that
$x\gets Sx$ and $y\gets Sy$. This has no effect on the inner product as $(Sx)^\top (Sy)=x^\top
S^\top S y=x^\top y$, since $S^\top S=I_n$. After the random rotation by $S$, $Sx$ becomes uniformly distributed over the sphere $\|x\|\mathbb{S}^{n-1}$,  and $Sy$ become
uniformly distributed on the sphere $\|y\|\mathbb{S}^{n-1}$, where $\mathbb{S}^{n-1}$ is the unit $n$-dimensional sphere. One can show that\footnote{To see this, first observe  $\EE\left[\frac{\|Sx\|^2_\infty}{\|Sx\|^2/n}\right]=\EE\|Z\|_{\infty}^2$ for $Z\sim\m{N}(0,I_n)$. Next, we use $\EE\|Z\|_{\infty}^2=\int_{0}^
\infty\Pr(\|Z\|_{\infty}^2> t)dt$ along with the upper bound $\Pr(\|Z\|_{\infty}^2> t)= 1-(1-2Q(\sqrt{t}))^n\leq\min\{1,n\sqrt{\frac{2}{\pi t}}e^{-t/2}\}$. For any $t_0>0$ we can therefore upper bound the integral by $t_0+n\int_{t_0}^\infty \sqrt{\frac{2}{\pi t}}e^{-t/2} dt\leq t_0+n\sqrt{\frac{8}{\pi t_0}}e^{-t_0/2}$. For all $n\geq 3$ we may take $t_0=2\ln n-\ln\ln n-\ln(\pi/2)>0$, and the resulting bound is smaller than $2\ln n$ for all $n\geq 27$.}  that as long as $n\geq 27$ 
\begin{align}
\EE[\Delta_{\mathrm{INT}}(x,y)]=\EE\left[\frac{\|Sx\|^2_\infty}{\|Sx\|^2/n}\right]\leq 2\ln n
~~  \forall x,y\in\RR^n\,.\nonumber
\end{align}

Applying random rotation for inducing an approximately Gaussian input to the quantizer, regardless of what the original data was,
has been a useful technique in lossy source coding and digital communication~\cite{popat1992robust,hung1998multidimensional,chen1998image,cai2000robust,feder2012method}. It has been recently applied to LLM post training quantization~\cite{chee2023quip,quarot2024,liu2024spinquant}. In order to avoid the computational complexity of rotation using Haar matrices, in practice a random Hadamard transform is typically used, or other low-complexity orthogonal~\cite{chee2023quip,ben2026quantizing} or nearly orthogonal transformations~\cite{popat1992robust}. 
We remark that another advantage of random rotation is that it reduces correlation between the data and quantization error. This was first demonstrated in~\cite{hadad2016dithered} and is used in federated learning
~\cite{vargaftik2022eden} and FP4-training ~\cite{chmiel2025fp4,panferov2025quest,panferov2026quartet}. 

Under random rotation, we obtain
\begin{align}
D_{\mathrm{INT}M,\mathrm{rotated}}\lesssim \frac{\|x\|^2\cdot\|y\|^2}{n} 2\cdot 2^{-2\Reff(\mathrm{INT}M))},
\label{eq:DintMrotated}
\end{align}
where $D_{\mathrm{INT}M,\mathrm{rotated}}$ is the \emph{expected} distortion and\footnote{In this paper rate is measured in bits, and therefore $\log$ is always taken to base $2$. We denote logarithm with the natural basis by $\ln$.} 
\begin{align}
\Reff(\mathrm{INT}M))=M-\frac{1}{2}\log \left(\frac{2\ln n}{3} \right).
\label{eq:ReffINT}
\end{align}
For example, for $n=4096$ we have that $\Reff(\mathrm{INT}M))\approx M-1.235$. In Section~\ref{subsec:numeric}, Table~\ref{tab:NormalizedError8} it is numerically verified that the approximation~\eqref{eq:DINTM} for the general case, as well as the approximation~\eqref{eq:DintMrotated} for the rotated case, are remarkably accurate. Note that the $\frac{1}{2}\log \left(\frac{2\ln n}{3} \right)$ rate penalty stems from using a single scale $\gamma_x,\gamma_y$ for each vector. If we use group-scaling, that is, allocate a different scale to each sub-block of size $m<n$, the rate penalty will be reduced to  $\frac{1}{2}\log \left(\frac{2\ln m}{3} \right)$. However, encoding the group-scales incurs a rate penalty of $c /m$ bits, where $c$ is the number of bits used for describing the scale. Thus, one needs to optimize $m$ in order to strike the right balance between the two effects.

For the Gaussian case where $x,y\sim\m{N}(0,\sigma^2 I_n)$ and are statistically independent, we have that $\EE\left[\frac{\|x\|^2\cdot\|y\|^2}{n}\right]=\sigma^4 n$ and that $\EE\left[\frac{\Delta_{\mathrm{INT}}(x,y)}{3} \right]\leq \frac{2}{3}\ln n$. While these quantities are not uncorrelated, for large $n$ their correlation vanishes, and we therefore treat them as uncorrelated.\footnote{A more careful inspection of $D_{\mathrm{INT}M}$ for the Gaussian case shows that the expression in~\eqref{eq:DintMGauss} is valid without any assumptions on the correlation between $\frac{\|x\|^2\cdot\|y\|^2}{n}$ and $\Delta_{\mathrm{INT}}(x,y)$.} We consequently obtain that for Gaussian iid $x,y$ and large $n$
\begin{align}
\frac{1}{2n\sigma^4}&\EE(\widehat{x^\top y}-x^\top y)^2 \nonumber\\
&\lesssim D_{\mathrm{INT}M,\mathrm{Gaussian}}\triangleq  2^{-2\Reff(\mathrm{INT}M))}.
\label{eq:DintMGauss}
\end{align}



\subsection{Floating Point Multipliers}

A signed floating point (FP) data-type consists of $1$ sign bit $s$, $\m{M}$ mantissa bits
$m_1,\ldots,m_{\m{M}}$ and $\m{E}$ exponent bits $e_1,\ldots,e_{\m{E}}$. It is further characterized by a fixed \emph{exponent-bias} term $\mu=2^{\m{E}-1}-1$. For example, the most popular FP8 format is E4M3, referring to $\m{E}=4$, $\m{M}=3$ and one sign bit. The constellation of points that can be represented by FP format contains the set
\begin{align*}
&\m{F}_{\mathrm{FP}}=\bigg\{(-1)^s\cdot 2^{E-\mu}\cdot\left(1+2^{-\m{M}}\cdot M \right)~:~s\in\{0,1\},\nonumber\\
&~E\in \{1,\ldots,2^{\m{E}}-2 \},M\in\{0,1,\ldots,2^{\m{M}}-1 \} \bigg\}\cup\{0\}.
\end{align*}
In fact, one can represent more numbers called subnormal numbers when $E=0$ and NaN and Inf when $E=2^{\m{E}}-1$, but for simplicity we omitted those options from the definition of $\m{F}_{\mathrm{FP}}$ above. For very small FP data-types such as FP4, these omitted values are important, and in~\eqref{eq:FP4const} we provide the full constealltion for FP4.

The FP quantization $Q_{\mathrm{FP}}(z)$ of a number $z\in\RR$ is as follows:
\begin{itemize}
\item Set $s_z\gets\Ind\{z<0\}$
    \item Set $E_z\gets\mu+\lfloor \log |z| \rfloor$, and $\bar{z}=2^{-(E_z-\mu)}\cdot |z|$. Note
    that $\bar{z} = 2^{-\lfloor \log |z| \rfloor}\cdot|z|\in[1,2)$.
    \item Set $M_z\gets\round\left(2^\m{M}\cdot (\bar{z}-1) \right)$
    \item If $M_z=2^\m{M}$, set $M_z\gets 0$ and $E_z\gets E_z+1$
\end{itemize}
We then set $Q_{\mathrm{FP}}(z)=(-1)^{s_z}\cdot 2^{E_z-\mu}\cdot\left(1+2^{-\m{M}}\cdot M_z \right)$. This is possible if $E_z\in\{1,\ldots,2^{\m{E}}-2\}$. Otherwise overload occurs, and we output $0$ if $E_z<1$ or the signed largest value in the constellation (if $E_z>2^{\m{E}}-1$).

We also define
\begin{align}
e_{\bar{z}}\triangleq 2^{-\m{M}}\cdot \round\left(2^{\m{M}}\cdot \bar{z}\right)-\bar{z},~\rho_z\triangleq \frac{2^{\lfloor \log |z| \rfloor}}{|z|}
\end{align}
and note that $\rho_z\in\left(\frac{1}{2},1\right]$, $e_{\bar{z}}\in2^{-\m{M}}\left[-\frac{1}{2},\frac{1}{2}\right)$, and that if overload did not occur we have that
\begin{align}
 Q_{\mathrm{FP}}(z)=z(1+\rho_z\cdot e_{\bar{z}}).\label{eq:idealizedFP}   
\end{align}
For the analysis that follows, we will adopt an approximate model which we call the
\emph{infinite-exponent independent noise} (IEIN) model. Under this model $Q_{\mathrm{FP}}(z)$ is
defined by equation~\eqref{eq:idealizedFP}, and $(e_{\bar{z}}, \rho_z,z)$ are mutually independent
with $e_{\bar{z}}\sim\Unif\left(2^{-\m{M}}\left[-\frac{1}{2},\frac{1}{2}\right) \right)$ and
$\rho_z=2^{-U}$ where $U\sim\Unif([0,1))$. Both approximations are justified by the observation
that while quantizing high-dimensional vectors, residuals of the rounding $\lceil\cdot\rfloor$ will
generally be equi-distributed in the $[-1/2,1/2)$ interval, as we discussed in
Section~\ref{sec:high-rate}.


The most straightforward way to approximate $x^\top y$ using FP multipliers is \emph{absmax quantization}. Below we describe \emph{dithered absmax quantization}, which is a simple variant of absmax quantization that is as easy to implement and will simplify the analysis:
\begin{itemize}
    \item Let $U_x,U_y\sim\Unif([0,1))$ be statistically independent. Let $\m{E}_{\mathrm{max}_{-}}=2^{\m{E}}-2-(\mu-1)$. Set
    $\gamma_x=2^{U_x}2^{-\m{E}_{\mathrm{max}_{-}}}\|x\|_{\infty}$, $\gamma_y=2^{U_y}2^{-\m{E}_{\mathrm{max}_{-}}}\|y\|_{\infty}$.
    \item Set $\tilde{x}=x/\gamma_x$ and $\tilde{y}=y/\gamma_y$
    \item Set $\hat{\tilde{x}}=Q_{\mathrm{FP}}(\tilde{x})$, $\hat{\tilde{y}}=Q_{\mathrm{FP}}(\tilde{y})$
    \item Set $\widehat{x^\top y}=\gamma_x\cdot\gamma_y\cdot \hat{\tilde{x}}^\top \hat{\tilde{y}}$.
\end{itemize}
When $U_x=U_y=1$ with probability $1$, this essentially reduces to standard absmax quantization.

Note that under dithered absmax quantization the IEIN model becomes quite realistic. In particular, 
\begin{align}
\rho_{\tilde{x}_i}=2^{\lfloor \log(\tilde{x}_i) \rfloor-\log(\tilde{x}_i)}=2^{-(\log(\bar{x}_i2^{-U_x})-\lfloor \log(\bar{x}_i2^{-U_x} \rfloor)}, \nonumber  
\end{align}
where $\bar{x}_i=2^{\m{E}_{\mathrm{max}_{-}}}x_i/\|x\|_{\infty}$. Since $U_x\sim\Unif([0,1)]$ is statistically independent of everything, we have that the IEIN assumption on  $\rho_{\tilde{x}_i}$ holds exactly for all $i\in[n]$. Similarly it holds for $\rho_{\tilde{y}_i}$ for all $i\in[n]$, and $\rho_{\tilde{x}_i}\indep \rho_{\tilde{y}_i}$ and are also independent of $x,y$. Furthermore, the scaling by $\gamma_x,\gamma_y$ prior to applying $Q_{\mathrm{FP}}(\cdot)$ guarantees that if overload did occur, it is only for entries whose magnitude is about $2^{-(2^{\m{E}}-3)}$ times smaller than the largest magnitude entry. Thus, if $n\ll 2^{-2\cdot (2^{\m{E}}-3)}$, the contribution of these entries to the total MSE in quantizing $\tilde{x}$ is negligible. The assumption that $e_{\bar{\tilde{x}}_i}$ is uniform is similar to the uniform quantization noise assumption we have made when analyzing INT$M$ constellations, and it becomes more accurate as the number of mantissa bits $\m{M}$ grows.

Under the IEIN model, the approximation error is 
\begin{align}
e_{\mathrm{FP}}&=\sum_{i=1}^n\gamma_x\gamma_y\hat{\tilde{x}}_i\hat{\tilde{y}}_i-x_i y_i\nonumber\\
&=\sum_{i=1}^n x_i y_i(1+\rho_{\tilde{x}_i}e_{\bar{\tilde{x}}_i})(1+\rho_{\tilde{y}_i}e_{\bar{\tilde{y}}_i})-x_i y_i\nonumber\\
&=\sum_{i=1}^n x_i y_i\left(\rho_{\tilde{x}_i}e_{\bar{\tilde{x}}_i}+\rho_{y_i}e_{\bar{\tilde{y}}_i} +\rho_{\tilde{x}_i}\rho_{\tilde{y}_i}e_{\bar{\tilde{x}}_i}e_{\bar{\tilde{y}}_i}\right).\nonumber
\end{align}
By our assumptions that $(e_{\bar{x}_i},e_{\bar{y}_i},\rho_{\tilde{x}_i},\rho_{\tilde{y}_i})$ are independent, and further assuming $\{e_{\bar{x}_i},e_{\bar{y}_i}\}$ are iid, we have
\begin{align}
\EE[e^2_{\mathrm{FP}}]=\frac{2^{-2\m{M}}}{12} \sum_{i=1}^n x^2_i y^2_i \left(2\ERho+(\ERho)^2\frac{2^{-2\m{M}}}{12} \right),
\label{eq:FPmseFull}
\end{align}
where 
\begin{align}
\ERho=\EE[2^{-2U}]\approx 0.541,~~~U\sim\Unif([0,1)]   
\end{align}
Neglecting the last term, we have
\begin{align}
&D_{\mathrm{FP},\m{M}}=\EE[e^2_{\mathrm{FP}}]\lesssim 2\ERho\cdot \frac{2^{-2\m{M}}}{12} \sum_{i=1}^n x^2_i y^2_i \nonumber\\
&~=  \frac{\|x\|^2\cdot\|y\|^2}{n} 2\cdot 2^{-2\Reff(\mathrm{FP},\m{M})}\cdot \Delta_{\mathrm{FP}}(x,y), \label{eq:DfpM}
\end{align}
where
\begin{align}
\Reff(\mathrm{FP},\m{M})=\m{M}+\frac{1}{2}\log\left(\frac{12}{\ERho}\right)\approx \m{M}+2.2356,  
\label{eq:ReffFP}
\end{align}
and
\begin{align}
\Delta_{\mathrm{FP}}(x,y)=    n\sum_{i=1}^n \frac{x_i^2} {\|x\|^2}\cdot\frac{y_i^2}{\|y\|^2}.\label{eq:DeltaFPdef}
\end{align}
In Section~\ref{subsec:numeric}, Table~\ref{tab:NormalizedError8} it is numerically verified that the approximation~\eqref{eq:DfpM} is remarkably accurate. 

Evidently, the gap from the fundamental limit is dictated by $\Delta_{\mathrm{FP}}(x,y)$. It is easy to see that $0\leq \Delta_{\mathrm{FP}}(x,y)\leq n$. If at least one of the vectors $x$ or $y$ is drawn from an iid distribution, independently of the other vector, it immediately follows that $\EE[\Delta_{\mathrm{FP}}(x,y)]=1$. Furthermore, if we apply a random rotation prior to quantization, such that $x\gets S x$ and $y\gets S y$, using the Cauchy-Schwartz inequality we have that
\begin{align}
\EE[\Delta_{\mathrm{FP}}(x,y)]&\leq n\sum_{i=1}^n \left(\EE\left(\frac{x_i}{\|x\|}\right)^4 \cdot \EE\left(\frac{y_i}{\|y\|}\right)^4\right)^{1/2}\nonumber\\
&=\frac{3n}{n+2}<3,    \nonumber
\end{align}
where we have used the fact~\cite[Appendix D]{ordentlich2024optimal} that if $U\sim\Unif(\alpha \mathbb{S}^{n-1})$, where $\mathbb{S}^{n-1}=\{z\in\RR^n~:~\|z\|=1\}$ is the unit sphere, then $\EE(U_1^4)=\frac{\alpha^4}{n^2}\frac{3n}{n+2}$.

\begin{figure}[t]
  \centering
  \includegraphics[width=\columnwidth]{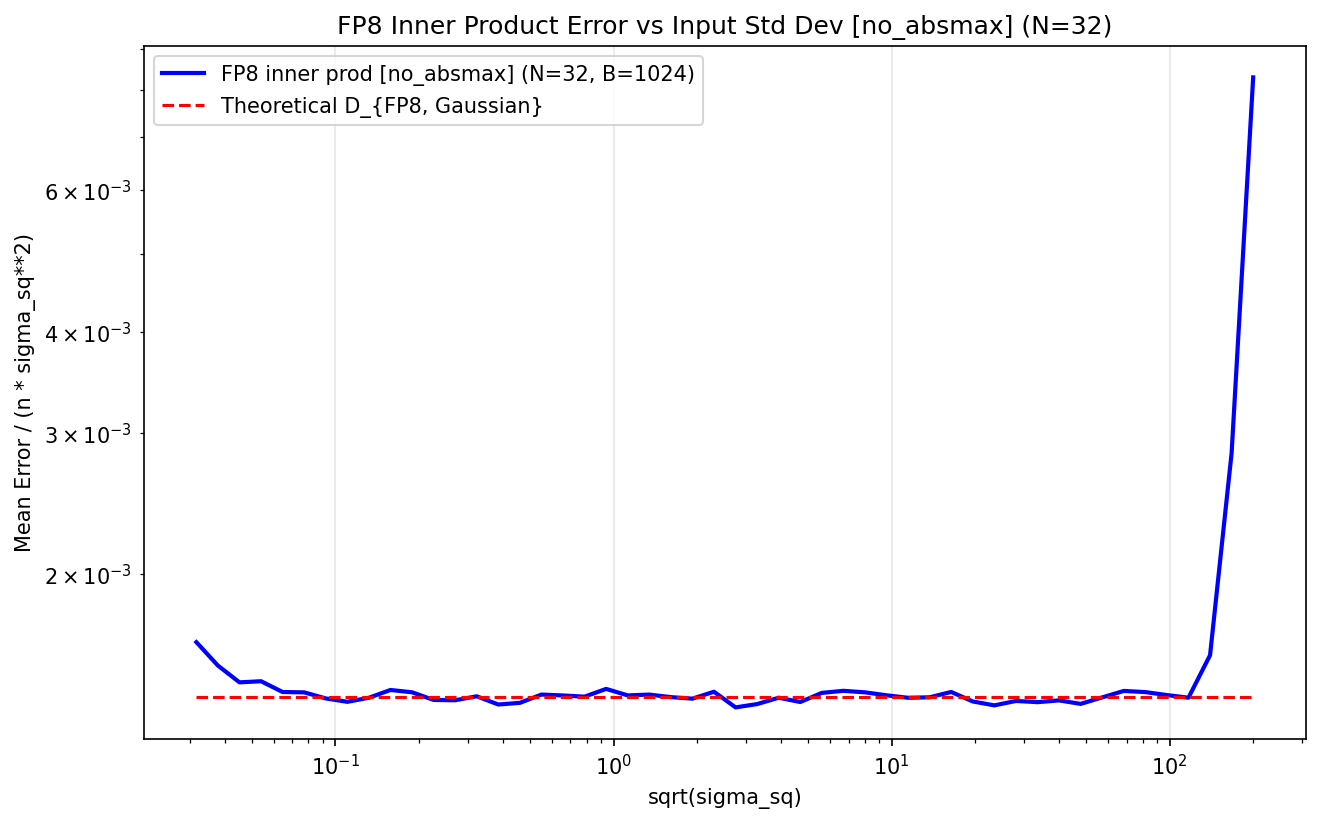}
  \caption{Demonstrating accuracy of FP8 approximation~\eqref{eq:DfpMGauss}. The figure plots
  average ratio (over 1024 pairs of vectors of dimension 32, each generated iid
  $\mathcal{N}(0,\sigma^2)$) of the normalized squared error of the inner-product (see left-hand
  side of~\eqref{eq:DfpMGauss}) against the simple theoretical approximation
  $2^{-2\Reff(\mathrm{FP},\m{M})) + 1}$. }
  \label{fig:fp8_sigma}
\end{figure}

For the Gaussian case where $x,y\sim\m{N}(0,\sigma^2 I_n)$ and are statistically independent, we
have $\EE\left[\frac{\|x\|^2\cdot\|y\|^2}{n}\right]=\sigma^4 n$ and $\EE\left[\Delta_{\mathrm{FP}}(x,y) \right]=1$ (this holds for any iid distribution
with zero-mean and variance $\sigma^2$). Furthermore, we have that $\frac{\|x\|^2\cdot\|y\|^2}{n}$ and $\Delta_{\mathrm{FP}}(x,y)$ are statistically independent. We consequently obtain that for Gaussian iid $x,y$ 
\begin{align}
{1\over 2n\sigma^4} &\EE(\widehat{x^\top y}-x^\top y)^2 \nonumber\\
&\lesssim D_{\mathrm{FP}\m{M},\mathrm{Gaussian}}\triangleq  
2^{-2\Reff(\mathrm{FP},\m{M}))}.
\label{eq:DfpMGauss}
\end{align}
It turns out that this approximation is accurate even without dithered absmax scaling for a wide range of
$\sigma$'s and even small dimensions $n$ (see
Fig.~\ref{fig:fp8_sigma}).

\begin{figure*}[t] 
    \centering
    
    \begin{subfigure}[b]{0.48\textwidth}
        \centering
        \includegraphics[width=\linewidth]{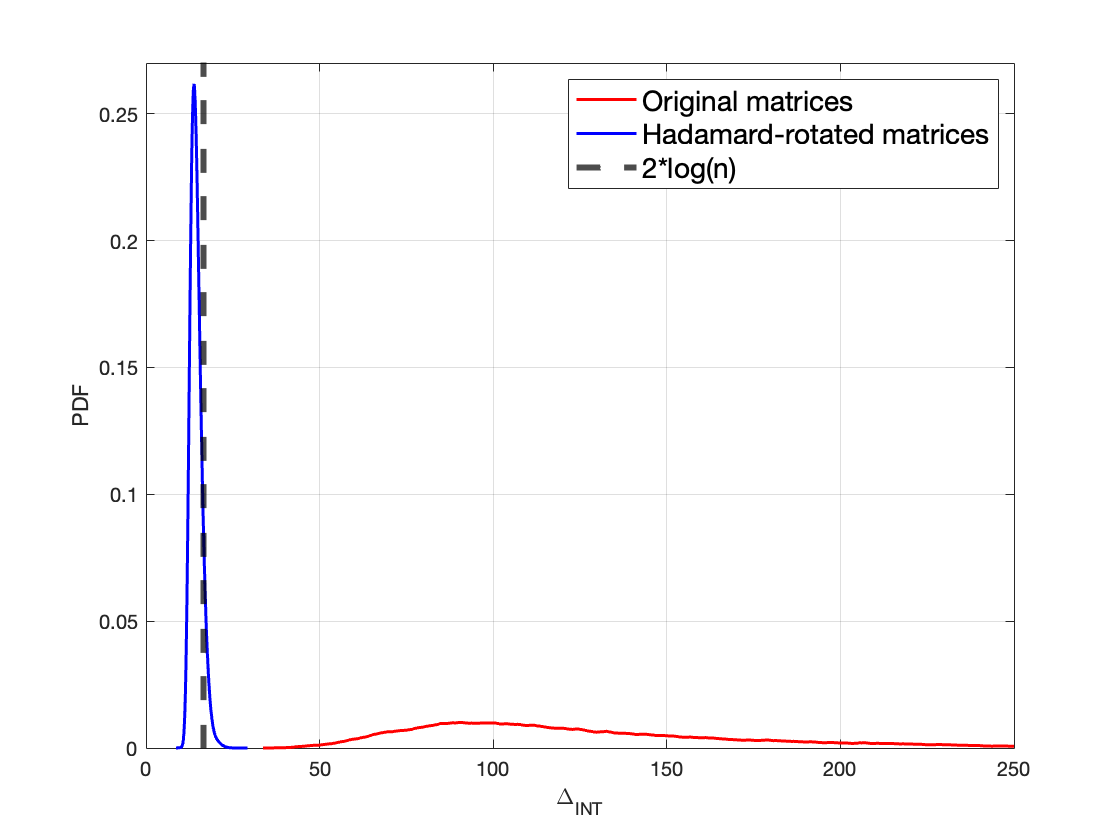}
        \caption{Histogram of $\Delta_{\mathrm{INT}}$}
        \label{fig:DeltaINT}
    \end{subfigure}
    \hfill 
    \begin{subfigure}[b]{0.48\textwidth}
        \centering
        \includegraphics[width=\linewidth]{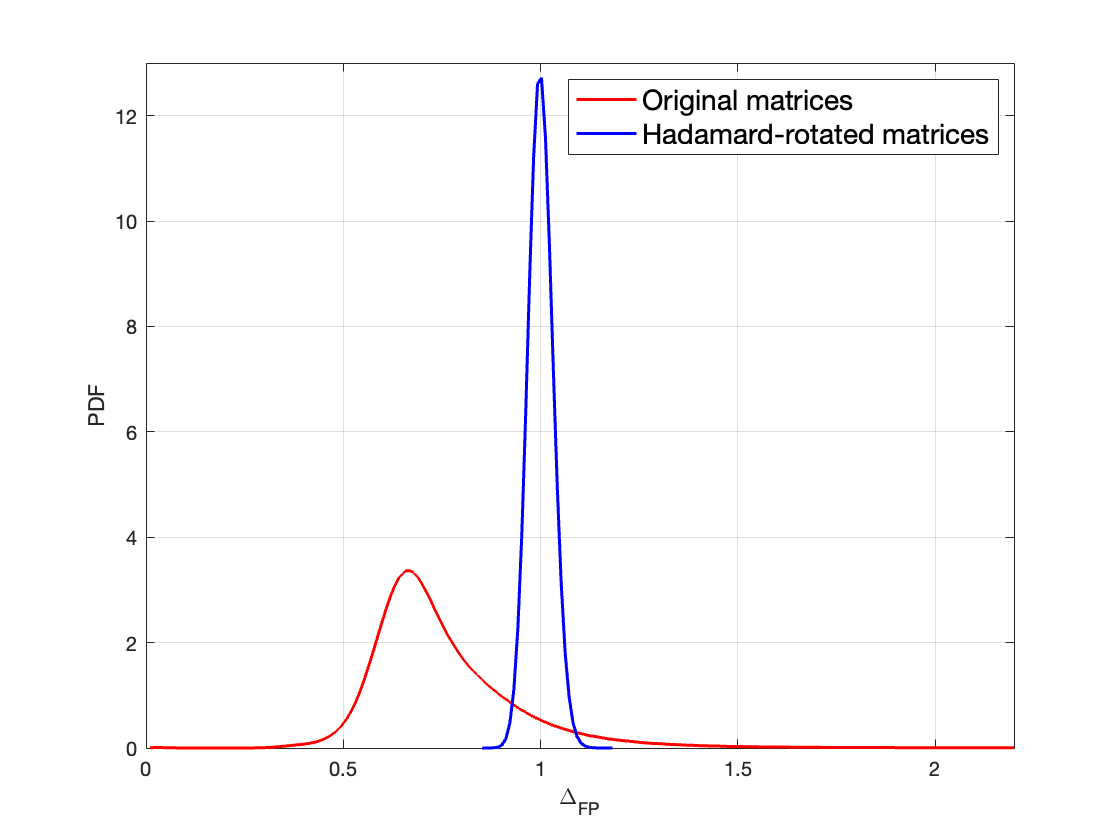}
        \caption{Histogram of $\Delta_{\mathrm{FP}}$}
        \label{fig:DeltaFP}
    \end{subfigure}
    
    \caption{Histograms of $\Delta_{\mathrm{FP}}$ and $\Delta_{\mathrm{INT}}$ for the setup
    described in Section~\ref{subsec:numeric}. }
    \label{fig:comparison}
\end{figure*}

    
    

\subsection{Multi-scaled INT and FP constellations}

A current trend in LLMs inference is to use $4$ bit multipliers whenever possible. Those include INT4 multipliers corresponding to entries in $\ZZ\cap[-8,7)$, and FP4 multipliers corresponding to entries in
\begin{align}
\m{FP}_4=\{0,\pm\frac{1}{2},\pm 1,\pm\frac{3}{2},\pm2,\pm4,\pm6\}.    
\label{eq:FP4const}
\end{align}
Recall that for INT$M$ multipliers with absmax scaling, the effective rate given by~\eqref{eq:ReffINT} is $\frac{1}{2}\log\left(\frac{2n}{3} \right)$ bits smaller than $M$, even when random rotation is applied. For $M$ as small as $4$, and even moderate $n$, using absmax scaling therefore decreases $\Reff(\mathrm{INT}M)$ significantly with respect to $M$. The FP4 constellation suffers from a similar problem: since it uses only $E=2$ exponent bits, overload errors are common under absmax scaling, and the attained distortion is far greater than that predicted by the IEIN model.

Consequently, those constellations are seldom used with absmax scaling. Instead, a multi-scaling
procedure is applied. Here we describe the NV multi-scaling mechanism: In order to compute $x^\top
y$, for $x,y\in\RR^n$, each vector is first scaled by a global scale, as in absmax scaling, but
afterwards it is split to consecutive sub-blocks of size 16. Then, each sub-block is scaled by an
E4M3 FP8 number, such that that the absolute value of the maximal entry in the scaled sub-block is
as close as possible to $7$ for INT4 or $6$ for FP8.\footnote{Alternatively, for each sub-block
one often chooses the minimal scale such that the largest value is at most $7$ (INT) or $6$ (FP). Another option is described in~\cite{cook2025four}.
Also, while NVFP4 is a popular data-type supported in Nvidia's modern GPUs, NVINT4 is far less
popular and is described here only for comparison purposes.} Consequently, both NVINT4 and NVFP4
actually have rate $4+\frac{8}{16}=4\frac{1}{2}$ bits per entry. 

We now make the heuristic argument that the distortion attained by NVFP4 is typically upper bounded by the distortion predicted by the IEIN model for $1$ mantissa bit (which has $\Reff(\mathrm{FP}\m{M})=3.2356$). Let us ignore the finite resolution of the E4M3 scale assigned to each vector $\tilde{x}\in\RR^{16}$, and assume for the analysis below that this scale is given in full resolution. Then, the input to the FP4 quantizer is $6\frac{\tilde{x}}{\|\tilde{x}\|_{\infty}}$. Assume without loss of generality that $\|\tilde{x}\|_{\infty}=|\tilde{x}_1|$, and define the set of coordinates $\m{OL}=\{i\in\{2,\ldots,16\}~:~6\frac{|\tilde{x}_i|}{\|\tilde{x}\|_{\infty}}<1\}$. For all the coordinates in $[16]\setminus\{1\cup\m{OL}\}$ there is no exponent overload, and the IEIN model is a good approximation. In contrast, for $i=1$ and $i\in\m{OL}$ the IEIN model is not a good approximation. For $i=1$, we have that $\tilde{x}_1$ is scaled exactly to a point $\pm6\in\m{FP}_{4}$, such that it suffers no quantization error. On the other hand, for all $i\in\m{OL}$, we have that the FP4 representation of $6\tilde{x}_i/\|\tilde{x}\|$ is in overload (its corresponding exponent is too small), and the IEIN model is overly optimistic. Since each one of these entries is quantized to either $\{0,\pm 1/2,\pm 1\}$, their total contribution to the distortion (after rescaling by $\|\tilde{x}\|_{\infty}/6$) is upper bounded by $|\m{OL}|\cdot\left(\frac{1}{4}\right)^2\left(\frac{\|\tilde{x}\|_{\infty}}{6}\right)^2$. Whenever $|\m{OL}|\leq 6< \frac{576}{48/\ERho}$, this is smaller than $\frac{1}{4}\frac{\|\tilde{x}\|^2_{\infty}}{12}\ERho$ which was ``saved'' (with respect to the IEIN model) by perfectly aligning the largest entry to the $\m{FP}_4$ constellation. Thus, for vectors in $\RR^{16}$ with $|\m{OL}|\leq 6$, the distortion for NVFP4 quantization is smaller than that predicted by the IEIN model with $\m{M}=1$.

\subsection{NestQuant}
\label{subsec:NQ}

The INT$M$ constellation admits very simple quantization and de-quantization procedures. However, it suffers from two shortcomings: 1)The distortion in quantization of a vector $x\in\RR^n$ depends on $\|x\|_{\infty}$ rather than on $\|x\|_2$; 2)The induced quantization cells are cubic, and therefore suffer from highly sub-optimal volume/second-moment tradeoff.\footnote{In fact, the strong slicing conjecture postulates that among all unit-volume convex sets at isotropic position, the cube has the largest second moment. See~\cite[Section 6.1]{regev2024reverse} and~\cite{klartag2025affirmative}} The NestQuant framework~\cite{ordentlich2024optimal,savkin2025nestquant}, which is the topic of this subsection, provides significant improvements with only a slight increase in the quantization/de-quantization complexity.

We review some basic lattice definitions. See~\cite{ramiBook} for a comprehensive treatment of lattices in information theory. For a lattice $L\subset\RR^d$ we define the nearest neighbor quantizer $Q_L:\RR^d\to L$ as
\begin{align}
Q_{L}(x)=\argmin_{\lambda\in L}\|x-\lambda\|, 
\end{align}
where ties are broken arbitrarily, but in systematic manner. The Voronoi region $\m{V}_L$ is defined as the set of all points in $\RR^d$ that are closer to $0$ than to any other lattice point
\begin{align}
\m{V}_L=\left\{x\in\RR^d~:~Q_L(x)=0\right\}.    
\end{align}
Any lattice $L\subset \RR^d$ has a (non-unique) generating matrix $G\in\RR^{d\times d}$ such that $L=G\ZZ^d$. The covolume of the lattice $L$, denoted $\mathrm{covol}(L)$, is the volume of its Voronoi region (or any other fundamental cell of $L$), which is also equal to $| G|$. The point density of a lattice is $\gamma(L)=\covol^{-1}(L)=|G|^{-1}$.
We define the second moment of the lattice $L$ as
\begin{align}
\sigma^2(L)=\frac{1}{d}\EE\|Z\|^2,    
\end{align}
where $Z\sim\Unif(\m{V}_L)$ is a random vector uniformly distributed over the Voronoi region of $L$. 

For a lattice $L\subset\RR^d$ and an integer $q\geq 2$ we have that $q L\subset L$ forms a \emph{self-similar} nested lattice pair. The lattice $q L$ is referred to as the \emph{coarse lattice} and it forms a partition of $L$ to $q^d$ cosets
\begin{align}
L=\bigcup_{i=1}^{q^d} \left(x_i+qL\right),\nonumber    
\end{align}
where $x_i\in L$, $i=1,\ldots,q^d$ are coset representatives. Note that any point in the coset $x_i+L$ can be chosen as the representative of this coset. Any choice of coset representatives induces a nested lattice codebook $L/qL$ consisting of $q^d$ points. A particularly useful choice is Voronoi codes~\cite{conway1982fast}, introduced by Conway and Sloane, where for each $i=1,\ldots,q^d$ the coset representative is chosen as the minimum energy member of this coset. In particular, the obtained codebook is of the form $L\cap q\m{V}_L$. Encoding of $x$ to $[q]^d$ amounts to computing $v=\mathrm{Enc}(x)=[G^{-1}Q_L(x)]\bmod q$, where here $\bmod ~ q$ denotes component-wise modulo $q$ reduction. Decoding amounts to computing $\mathrm{Dec}(v)=Gv-q\cdot Q_L(Gv/q)$.\footnote{We ignore dithering here, for simplicity, though it is needed for preventing undesired boundary effects~\cite[Section II]{ko25}.} Thus, if we have an efficient implementation of the nearest neighbor lattice quantizer $Q_L(\cdot)$, we can also implement the encoder and decoder of Voronoi codes efficiently. Notably, \emph{the complexity of encoding and decoding does not grow with $q$} (thus, high-rate and low-rate quantization have the same computational cost). Furthermore, whenever $Q_L(x)\in q\m{V}_L$ we have that $\mathrm{Enc}(\mathrm{Dec}(x))=Q_L(x)$, and the quantization error is in $\m{V}_L$. We refer to the event $Q_L(x)\notin q\m{V}_L$ as an \emph{overload} event.

Note that the INT$M$ constellation is in fact a Voronoi code for the nested lattice pair $(\beta\ZZ^n/2^M\beta\ZZ^n)$ for $\beta>0$. In order to avoid overload when quantizing $x\in\RR^n$, the absmax scaling sets $\beta=2^{-(M-1)}\|x\|_{\infty}$, and the corresponding second moment is $\beta^2\sigma^2(\ZZ)$, where $\sigma^2(\ZZ)=\frac{1}{12}$.

NestQuant applies a random rotation $S$ (in practice, randomized Hadamard transform) to both vectors $x,y\in\RR^n$ prior to quantization. Then, it splits each rotated vector to chunks of size $d$, and quantizes each one of them separately. It improves over INT$M$ by using:
\begin{enumerate}
\item \textbf{Better lattice:} Replacing $\ZZ^d$ with a lattice $L\subset \RR^d$ of the same covolume, that admits efficient nearest neighbor decoding $Q_{L}(\cdot)$ and has smaller second moment $\sigma^2(L)<\sigma^2(\ZZ)=1/12$ and smaller overload probability $$\Pr(Q_{\beta L}(X)\notin q\beta\m{V}_L)<\Pr(Q_{\beta \ZZ^d}(X)\notin q\beta\m{V}_{\ZZ^d}),$$ for $X\sim\m{N}(0,I_d)$;
\item \textbf{Multi-scaling:} Using a bank of $K$ different scales $\{\beta_1,\ldots,\beta_K\}$ and using the scale that results in the smallest squared error. The effective codebook is therefore
\begin{align}
\m{C}=\bigcup_{k=1}^K \beta_k( L\cap q\m{V}_L).\nonumber
\end{align}
\end{enumerate}
In~\cite{ordentlich2024optimal} it was shown that for $d=n$ with $n$ large enough, there exist
lattices $L$ for which NestQuant attains the optimal distortion-rate tradeoff for matrix
multiplication. However, such lattices typically do not admit efficient $Q_{L}(\cdot)$.
In~\cite{savkin2025nestquant} it was shown that by using $d=8$ with the lattice $L=E_8$, which has
a very fast $Q_{L}(\cdot)$, one obtains excellent performance, significantly outperforming other
LLM quantization strategies at a similar bit-rate.

\subsection{Numerical results}
\label{subsec:numeric}

The goal of this section is to give numerical validation to our analysis of INT and FP constellations, which relied on some approximations, and compare performance of various quantizers in a realistic setup. To this end we focus on one particular linear operation within an LLM, and consider the $W_v\in\RR^{n\times a}$ matrix in the $15$th layer of Llama3-8B, along with its corresponding activation vectors. Here, $n=4096$, $a=1024$, and the matrix $X\in\RR^{b\times n}$ consists of $b=10,000$ rows, which were sampled by running 5 different prompts on the model and taking about $2048$ consecutive activation vectors for each prompt.

For these matrices we define $3$ matrices of size $b\times a$:
\begin{align}
K(i,j)&=2\frac{\|X(i,:)\|^2\|W(:,j)\|^2}{n}~~i\in[b],j\in[a]\nonumber\\
\Delta_{\mathrm{INT}}(i,j)&=\Delta_{\mathrm{INT}}(X(i,:),W(:,j))~~i\in[b],j\in[a]\nonumber\\
\Delta_{\mathrm{FP}}(i,j)&=\Delta_{\mathrm{FP}}(X(i,:),W(:,j))~~i\in[b],j\in[a]\nonumber,   
\end{align}
where $\Delta_{\mathrm{INT}}$ and $\Delta_{\mathrm{FP}}$ are defined in~\eqref{eq:DeltaINTdef} and~\eqref{eq:DeltaFPdef}, respectively. 

First, we explore the accuracy of our approximations for $D_{\mathrm{INT}M}$ and $D_{\mathrm{FP}\m{M}}$, as given in~\eqref{eq:DINTM} and~\eqref{eq:DfpM}. To this end we quantize $X$ and $W$ using INT8 absmax quantization and using FP8 (E4M3) dithered absmax quantization. We denote the approximation error of $XW$ using these methods by $e_{\mathrm{INT}8}\in\RR^{b\times a}$ and $e_{\mathrm{FP}8}\in\RR^{b\times a}$, respectively. Table~\ref{tab:NormalizedError8} provides the root mean squared error (RMSE) of the $ab$ error entries with and without Hadamard rotation, under different normalizations. We denote
\begin{align}
\bar{e}_{\mathrm{INT}8}(i,j)&=\frac{e_{\mathrm{INT}8}(i,j)}{\sqrt{K(i,j)\Delta_{\mathrm{INT}}(i,j)/3}},~i\in[b],j\in[a]\nonumber\\
\bar{e}_{\mathrm{FP}8}(i,j)&=\frac{e_{\mathrm{FP}8}(i,j)}{\sqrt{K(i,j)\Delta_{\mathrm{FP}}(i,j)}}~i\in[b],j\in[a].\nonumber
\end{align}
If our expressions for $D_{\mathrm{INT}M}$ and $D_{\mathrm{FP}\m{M}}$ are accurate, then for $I\sim\Unif([b])$ and $J\sim\Unif([a])$ it holds that
\begin{align}
\EE(\bar{e}^2_{\mathrm{INT}8}(I,J))=2^{-2M},~\EE(\bar{e}^2_{\mathrm{FP}8}(I,J))=2^{-2\Reff(\mathrm{FP})}.\nonumber
\end{align}
The first row of Table~\ref{tab:NormalizedError8} shows that those approximations are indeed
remarkably accurate. 

The second row of Table~\ref{tab:NormalizedError8} provides the RMS of the quantization errors normalized by $\sqrt{K(i,j)}$. Recall from~\eqref{eq:fundlimit} that the fundamental limit for generic rate $R$ quantized MatMul is $D_{ij}=K(i,j)\cdot 2^{-2R}$, assuming $R\gg1$. Thus, if the optimal high-dimensional rate $R=8$ quantizers from~\cite{ordentlich2024optimal} were used instead of the various INT8/FP8 schemes investigated, the normalized RMSE would have been $2^{-8}$. The second row in Table~\ref{tab:NormalizedError8} therefore shows how many of the 8 bits each of the scheme is using are ``effective''.

The third row of Table~\ref{tab:NormalizedError8} provides the RMS of the quantization errors normalized by $\sqrt{2n}$ for the case where $X,W$ are of the same shape as above, but their entries are drawn iid $\m{N}(0,1)$. The results verify the accuracy of our expressions for $D_{\mathrm{INT}\m{M},\mathrm{Gaussian}}$ and $D_{\mathrm{FP}\m{M},\mathrm{Gaussian}}$ in~\eqref{eq:DintMGauss} and~\eqref{eq:DfpMGauss}, respectively. Note that, as expected, the effective rate for rotated INT$M$ quantization is the same as that of INT$M$ quantization for Gaussian matrices.

In order to obtain a deeper understanding of where the numbers in the second row in Table~\ref{tab:NormalizedError8} come from, we explore the distribution of $\{\Delta_{\mathrm{INT}}(i,j)\}_{i\in[b],j\in[a]}$ and $\{\Delta_{\mathrm{FP}}(i,j)\}_{i\in[b],j\in[a]}$.
In Figure~\ref{fig:DeltaINT} we plot the histogram of the $a\cdot b$ entries of the matrix $\Delta_{\mathrm{INT}}$ with and without applying Hadamard rotation on both matrices. We see that for the original matrices $\Delta_{\mathrm{INT}}$ can take very large values, whereas for Hadamard rotated matrices the entries of $\Delta_{\mathrm{INT}}$ concentrate below $2\ln(n)$. Thus, rotating the matrices prior to INT quantization is crucial (at least for Llama3-8B).

In Figure~\ref{fig:DeltaFP} we plot the histogram of the $a\cdot b$ entries of the matrix $\Delta_{\mathrm{FP}}$ with and without applying Hadamard rotation on both matrices. Here, the situation is very different. While Hadamard rotation makes the distribution concentrated around $1$, we see that without rotation the entries of $\Delta_{\mathrm{FP}}$ are typically smaller than $1$. This shows that not only is full-vector rotation unnecessary for FP quantization, it is actually harmful! Of course, this conclusion is specific for the matrices we used for this experiment. However, prior work~\cite[Table 1]{chen2025wush},~\cite{shao2025block} have already observed that rotation is often harmful for FP LLM quantization, and we argue that the behavior of $\Delta_{\mathrm{FP}}$ provides a rigorous explanation for this phenomenon.

\begin{table*}[h!]
    \centering
    \renewcommand{\arraystretch}{1.5} 
    \begin{tabular}{|l|c|c|c|c|} 
        \hline
        \textbf{} & \textbf{INT8} & \textbf{Hadamard INT8} & \textbf{FP8} & \textbf{Hadamard FP8} \\ 
        \hline
        $\mathrm{RMS}\left(\bar{e}_{\mathrm{INT}8/\mathrm{FP}8}(i,j)\right)$ & $2^{-7.9994}$ & $2^{-8.0012}$ & $2^{-5.2346}$ & $2^{-5.2372}$ \\ 
        \hline
        $\mathrm{RMS}\left(\frac{e_{\mathrm{INT}8/\mathrm{FP}8}(i,j)}{\sqrt{K(i,j)}}\right)$ & $2^{-5.2495}$ & $2^{-6.8664}$ & $2^{-5.4378}$ & $2^{-5.2370}$ \\ 
        \hline
        $\mathrm{RMS}\left(\frac{e_{\mathrm{INT}8/\mathrm{FP}8}(i,j)}{\sqrt{2n}}\right)$ iid $\m{N}(0,1)$ entries & $2^{-6.8619}$ & $2^{-6.8645}$ & $2^{-5.2395}$ & $2^{-5.2383}$ \\ 
        \hline
    \end{tabular}
    \caption{RMS of normalized errors}
    \label{tab:NormalizedError8}
\end{table*}

\begin{figure}[t]
  \centering
  \includegraphics[width=\columnwidth]{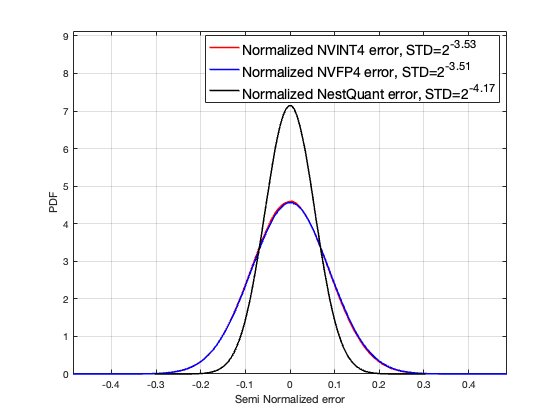}
  \caption{Performance of several quantization schemes with $R=4.5$.}
  \label{fig:NVRD}
\end{figure}

Figure~\ref{fig:NVRD} compares the approximation error distribution of $3$ quantization schemes with rate $R=4.5$ bits per entry: NVINT4 after Hadamard rotation, NVFP4 without Hadamard rotation, and NestQuant with $L=E_8$ and a bank of $K=16$ different scales. For all schemes, we normalize the approximation error of the $ij$th entry by $\sqrt{K(i,j)}$. The performance of NVINT4 and NVFP4 is seen be quite similar, while that of NestQuant is significantly better and attains $\approx 0.6$ more bits of accuracy with respect to the former schemes.

\section*{Acknowledgment}
We thank Egor Lifar and Semyon Savkin (MIT EECS) for numerous discussions
and help with obtaining calibration matrices for Llama-3-8B. We also thank Uri Erez (TAU EE) for helpful suggestions.

\bibliographystyle{IEEEtran}
\bibliography{IPRD_Bib}

\end{document}